# CW frequency doubling of 1029 nm radiation
# Using single pass bulk and waveguide PPLN crystals

*N. Chiodo*[a], *F. Du Burck*[b], *J. Hrabina*[c], *Y. Candela*[a], *J.-P. Wallerand*[d] *and O. Acef*[a]

[a] LNE-SYRTE / Observatoire de Paris/ CNRS UMR 8630 / UPMC Paris VI,
61, avenue de l'Observatoire, 75014 Paris, France

[b] Laboratoire de Physique des Lasers, Université Paris 13, Sorbonne Paris Cité, CNRS,
99 Avenue Jean-Baptiste Clément, 93430 Villetaneuse, France

[c] Institute of Scientific Instruments of the ASCR, v.v.i ,
Královopolská 147, 612 64 Brno, Czech Republic

[d]: Laboratoire commun de métrologie, LNE-CNAM
1, rue Gaston Boissier, 75724 Paris Cedex 15

* Corresponding author: nicola.chiodo@obspm.fr

**Abstract**

Following various works on second harmonic process using periodically poled Lithium Niobate crystals (PPLN), we report on the performances comparison between commercial bulk and waveguide crystals at 1029 nm. We use a continuous wave (CW) amplified Yb doped single fibre laser delivering up to 500mW in single mode regime. In case of bulk crystal we generate 4 mW using 400 mW IR power. The use of waveguide crystal leads to an increase of the harmonic power up to 33mW with input IR power limited to 200mW. Nevertheless, this impressive efficiency was affected by the long term degradation of the non-linear waveguide crystal.

## 1. Introduction

In recent years a great deal of effort has focused on the development of ultrastable and powerful lasers based on very compact experimental setups. These laser sources are of considerable interest for space and/or terrestrial applications ranging fundamental physics tests, frequency metrology, coherent optical communications, high resolution spectroscopy, ultra stable optical clock, etc, [1-6]. Beside development and use of rigid optical cavities which offer the possibility to confer impressive frequency stability to IR lasers, atomic and molecular lines permit more simple and less sensitive to seismic and thermal environment. Moreover, they constitute absolute frequency references. For this frequency stabilisation purpose, molecular or atomic resonances are required to be strong and very narrow at same time. Such lines exist in the visible range, with high quality factor and large signal to noise ratio [7,8]. In contrast, in the IR domain several atomic or molecular species are available in coincidence with laser emission spectrum, but they do not meet both requirements in the same time. On the other hand, thanks to their noticeable compactness, laser diodes emitting in the visible domain could be the best candidates as laser sources for these developments, unfortunately they suffer from their low power and intrinsic large linewidth. On the contrary, solid state lasers such as Nd:YAG or Yb:YAG, or fiber lasers, emitting in the IR offer attracting possibilities regarding to their low intrinsic phase noise, strong output powers and robustness.

Non linear crystals permit to bridge in simple way the IR–to-visible ranges of the optical domain. In this way, achievement of strong harmonic powers generation gives the possibility of interrogation of several molecular resonances for efficient frequency stabilization of the IR lasers. Due to its very high nonlinear properties, the Lithium Niobate (LN) [9] is one of the most important crystals for harmonic generation and frequency mixing in a large domain from IR to visible in a basic single-pass and very compact configuration. The advent of periodically poled structures (PP) during the last decade permits easy phase matching and exploitation of the highest nonlinear coefficient $d_{33}$ for impressive efficient frequency conversion [10]. This conversion efficiency can be even more enhanced by at least one order of magnitude using a waveguide structure [11,12]. For that purpose, a precise positioning adjustment of the crystal with respect to the IR beam is required in order to maximize the coupled IR power in the waveguide. Nevertheless, these constraints are still much easy to be satisfied compared to intra-cavity frequency doubling when the targeted harmonic powers are below Watt level [13]. The main limitation relies to the optical damage due the extremely high optical density in the waveguide.

In this paper we present a comparative study on the performances and characteristics of PPLN bulk and waveguide crystals for CW doubling at 1029 nm in single pass configuration. Waveguide crystals show efficiencies greater than $100\%.cm^{-1}.W^{-2}$. At higher power, an optical conversion efficiency of 17% was obtained with one of them, giving the possibility to achieve a power of 33 mW of green from 197 mW of incident IR power, with a power coupling in the waveguide of 45%. However, for such input power a deterioration of several tens % of the efficiency was observed after ten hours of irradiation. The initial efficiency was restored by stopping the irradiation during several days. On the long term, anyway the conversion efficiency of waveguide crystals has been progressively reduced in a permanent way, and after more than 2 years of use, this value is no more than 60% of the initial one.

## 2. Crystals for SHG

The SHG is performed using 3 different MgO:PPLN commercial crystals in a single pass configuration: one of them (crystal A) is bulk and the others (crystal B and C) are waveguide crystals. All the crystals are manufactured to generate second harmonic (SH) radiation at 515 nm.

| Crystal | Substrate (mm) | | | Guide (µm) | | Poling Period (µm) |
|---|---|---|---|---|---|---|
| | Length | Width | Height | Width | Height | |
| A | 25 | 2 | 0.5 | no | no | 6.10 |
| B | 10 | 1.8 | 0.5 | 5 | 6 | 6.07 |
| C | 10 | 1.8 | 0.5 | 3 | 6 | 6.06 |

*Tab. 1: Geometrical parameters and poling periods of the PPLN crystals used in the experiment*

The geometrical parameters of the three crystals used in this work are reported in Tab. 1. In case of crystals B and C, the waveguides are produced by ion-exchange technique 2 µm below the surface of the crystal and are designed to generate the second harmonic of 1029 nm with a quasi-phase matching (QPM) temperature around 50°C. The input and output facets of the waveguide are AR coated at both 1030 nm and 515 nm. In case of crystal C, which has the smallest transversal dimensions, the numerical aperture (N.A.) of the waveguide is 0.16.

**3. Experimental setup**

The laser source is an Ytterbium-doped DFB fibre laser that delivers 12mW near 1029 nm amplified up to 500mW. The laser linewidth is lower than 10 kHz. The frequency of the laser can be adjusted by means of temperature control of the Bragg grating or with a ceramic piezo-electric actuator (PZT) which stresses this Bragg grating. Therefore, the laser wavelength can be tuned from 1028.7 to 1029.4 nm. A linear polarization is defined before the optical amplifier. The laser system is placed in a metallic box which is enclosed in an additional wood box to ensure good thermo-acoustic isolation. This packaging limits the long term free running frequency drift to less than 50 MHz in 2 days. A half wavelength plate is located at the laser output to adjust the direction of the linear polarization.

The experimental setup is depicted in Fig. 1. For measurements with crystals B and C, an acousto-optics modulator (AOM) is inserted at the output of the laser. It controls the power sent towards the doubling crystal and can be used in a servo-loop for controlling the SH power, as described in section 5.2.

At the output of the AOM, the linear polarization is adjusted with a half-wave plate and lens $L_4$ is used to couple the power in the crystal, which focal length is adapted to the used crystal. A dichroic mirror separates the fundamental frequency beam from the second harmonic one (green beam). A part of the green beam is taken by an adjustable beamsplitter (half wave plate and $PBS_2$) and monitored by a photodiode (PD). It is used to stabilize the IR power at the crystal input by a servo loop driving the RF AOM power.

Each crystal is inserted in a temperature-controlled oven as shown in figure 2, yielding to a stabilized temperature within less than 5 mK during several tens hours (Fig. 3). The fast oscillations in Fig. 3 are due to the air-conditioning system of the laboratory while the long term fluctuations correspond to the daily environmental variations. These residual temperature fluctuations are at least 3 orders of magnitude lower than the temperature acceptance of all the crystals studied in this work.

**4. Bulk crystal**

The laser beam is focused inside the crystal, with lens L4 (focal length = 10 cm). The SHG power versus the input wavelength is shown in Fig. 4(a). The crystal is characterized at 31.0°C with an input IR power of 400 mW, limited by the power available at the laser output. In these conditions, a maximum SHG of 4 mW is found at 1029.35 nm. The fit of these data using a $sinc^2$ results in a wavelength acceptance of 0.1 nm. The SH power versus the temperature of the crystal is plotted in Fig. 4(b). The temperature acceptance deduced from a fit of the data shown in fig. 4(b) is 1°C, much higher than our stabilization capability.

In fig. 4(c), the SHG power is plotted versus the square of the IR power at 1029.35 nm. The data show a very good agreement with the quadratic model expected for undepleted frequency doubling. The slope of the line is 2.55% $W^{-1}$ corresponding to 1% $W^{-1}.cm^{-1}$.

**5. Waveguide crystals**

*5.1. SH efficiency and evolution in time*

The transverse dimensions of channels are 6 µm x 5 µm for crystal B and 6µm x 3µm for crystal C. Both crystals are inserted on the same mount at the same height, which gives the possibility of an easy changing of the working crystal with a simple translation (Fig. 2).

For both crystals, the higher harmonic power was obtained using lens L4 with focal length of 8 mm, NA of 0.5. The coupling efficiency in the waveguide is measured in the following way. In order to avoid errors due to the converted power, measurements must not be operated in phase matching conditions. We first search the maximum of the SH power for given temperature and wavelength, then we adjust the temperature to reach the first zero of the SH power. Besides, in order to avoid detection of the uncoupled light, the IR power meter is placed at 40-60 cm from the output of the crystal after lens $L_5$ and a diaphragm is inserted in the path. An IR coated lens is used to focus the light on the power meter. The coupling efficiency is measured with an input IR power of 135 mW. It is found to be 63% for crystal B and 45% for crystal C. Those values are significantly smaller than the values of about 80% given by the manufacturer, but it must be emphasized that those latter ones are achieved by a direct fibre coupling at the crystal input [12,14].

The waveguide crystals have been used for more than 2 years and we have observed a progressive, but permanent degradation of the performances of SHG process. Fig. 5 shows the SH power $P_{2\omega}$ measured at the output of the crystal for both crystals. The filled symbols correspond to data recorded at the beginning of their use. For those measurements, the SH power was first optimized at the maximum IR power for a given wavelength by optimizing both the IR power coupling and the crystal temperature. Then the measurements are operated by decreasing the input IR power. For each point, only the IR power coupling was adjusted to optimize the SH power, the temperature remaining the same. The solid curves in Fig. 5 are the theoretical evolution of the SH power versus the input IR power given by

$$P_{2\omega} = \tau P_\omega \tanh^2\left(\sqrt{\eta L^2 \tau P_\omega}\right) \quad (1)$$

where $\tau$ is the coupling efficiency ($\tau = 63\%$ for crystal B and $\tau = 45\%$ for crystal C) at the waveguide input and $L = 1$ cm is the waveguide length. The normalized conversion efficiency $\eta$ determined from low power measurements is found to be $\eta = 278\% \cdot W^{-1} \cdot cm^{-2}$ for crystal B and $\eta = 613\% \cdot W^{-1} \cdot cm^{-2}$ for crystal C. It is seen that both crystals show the same trend in good agreement with the theoretical model given by (1). However the SH power at the output of crystal B clearly departs from the theoretical curve above 150 mW of IR input power and some saturation appears (dash line).

This effect was already reported for waveguide PPLN crystals [15-17]. It was shown to be essentially due to the green induced infrared absorption (GRIIRA) in the guide [17]. The optical conversion efficiency $\eta_{opt} = 100 \times P_{2\omega}/P_\omega$ versus the IR power $P_\omega$ at the input of the crystal is calculated for both crystals and plotted in Fig. 6. The solid curves are derived from the values of normalized conversion efficiencies $\eta$ deduced from Fig. 5. The maximum green power obtained at the output of crystal B was 30.5mW with a pump power of 232mW at 1029.3nm which corresponds to an optical conversion efficiency of 13%. For crystal C, the maximum output green power was 33.3mW with a IR power of 197mW, corresponding to an optical conversion efficiency of 17%.

To be noted that the nominal efficiency of the crystal B is about half that of crystal C (80%W$^{-1}$cm$^{-2}$ and 200%W$^{-1}$cm$^{-2}$, respectively), but the SH generated power for both crystals is almost the same due to the less efficient coupling of crystal C.

The stability in time of the conversion efficiency of crystal C has been studied. The crystal was coupled with 180mW of IR power (81mW coupled in the waveguide) during 12 hours. The initial SH output power of about 29mW fell down to 21.5mW at the end of this period, corresponding to a decrease of about 26%. A new irradiation with the same input power during 12 hours leads to a new decrease and only 50 % of the initial green power was available (15mW). It was possible to re-obtain the initial 29mW of output green power from 180mW of IR input power by leaving the crystal at rest during 10 days (the IR is blocked). In a second step, we performed the same test with an input power divided by three. The experiment is conducted with an input IR power of 60mW (27mW coupled in the waveguide) leading to 4mW of green output power. In this case, no degradation was noted after 12 hours of irradiation.

Nevertheless, the SHG efficiency conversion for both crystals has gradually degraded in a permanent way after 2 years of use as illustrated by Fig. 5 in which the empty symbols correspond to data recorded recently. For each measurement, the IR power coupling and the crystal temperature are optimised. The solid curves correspond to the theoretical evolution of the SH power given by eq. (1).

The value of the normalized conversion efficiency $\eta$ deduced from the low power measurements is found to be $\eta = 182\% \cdot W^{-1} \cdot cm^{-2}$ for crystal B and $\eta = 357\% \cdot W^{-1} \cdot cm^{-2}$ for crystal C. One observes a decrease of the efficiency by a factor 1.5 for crystal B and by a factor 1.7 for crystal C.

The optical conversion efficiency $\eta_{opt} = 100 \times P_{2\omega}/P_\omega$ versus the IR input power $P_\omega$ is calculated in each case and plotted in Fig. 6. The solid curves are deduced from the values of normalized conversion efficiencies $\eta$ of Fig. 5. The maximum optical conversion efficiency $\eta_{opt}$ is found to be roughly 8% for both crystals.

The degradation of SHG efficiencies is not associated with a decrease of the IR coupling efficiencies in the crystals which remained the same from the beginning of their use. We have also verified that there was no degradation of the AR coating of the input face of the crystal by reversing the direction of propagation in the guide and using the output face as input one. At last, let us note that the SH power dependence with temperature and laser wavelength, detailed in the section 5.2, did not change from the beginning of the use of the crystals.

*5.2. Other characteristics of waveguide crystals*

Fig. 7 shows the wavelength dependence of the SHG power, for the 2 waveguide crystals, collected at fixed temperature and fixed IR power. Due to the limited wavelength range allowed by the laser, it was not possible to measure the complete curve for crystal B whereas this was possible for crystal C. For crystal B, the temperature was first optimised to obtain the maximum SHG at the limit of the laser tuning range and the laser wavelength was then decreased maintaining the crystal input IR power constant. The data of Fig. 7 are fitted with a sinc$^2$ function to determine the width of the curves. The wavelength acceptance is found to be 0.25nm for crystal B and 0.2nm for crystal C. The SHG power versus the crystal temperature is shown in Fig. 8 for both crystals. Temperature acceptance is 2.47 °C for crystal B and 2.42°C for crystal C.

In Fig. 7 and Fig. 8, one can observe for both crystals some discrepancies between the experimental data and the theoretical sinc$^2$ shape in the lateral lobes. For crystal B, the SH power does not exactly vanish as predicted by theory. This behavior has already been reported in ref. [15,18] and may be related to inhomogeneities of the waveguide which prevent the destructive interference of the SH power.

In case of crystal C, a strong asymmetry is observed between upper and lower lobes. This effect can be attributed to inhomogeneities of the polling period or index of the guide [17,19].

The evolution of the QPM temperature with the coupled IR power has been characterized for both crystals. It is found that the QPM temperature slightly decreases with the IR power injected in the guide, while the shape of the curve of the SH power as a function of the crystal temperature remains the same, as shown in Fig. 9 for crystal B. The inset of Fig. 9 shows the QPM temperature versus the coupled IR power. The linear fit of the data gives a slope of -9.7mK/mW for crystal B and -10.4mK/mW for crystal C. This behavior is reported in literature with various magnitudes [12,14,16,17]. To understand this behavior, it must be emphasized that one does not directly measure the temperature of the waveguide, but the temperature of the crystal mount to which the temperature sensor is attached. Similary, the servo loop adjusts the temperature of the crystal mount. If a part of the optical power is absorbed in the guide (absorption of the medium, GRIIRA, etc), a heat source appears and the temperature of the guide becomes higher than that of the mount. This temperature shift increases with absorption. To ensure the QPM, one must decrease the setpoint temperature of the servo loop, as seen in Fig. 9. The magnitude of this effect depends of the thermal couplings and may vary from one setup to another.

The stabilization of the green power in the long term was performed for crystal C. The green power was monitored and the IR input power was driven by the AOM located before the crystal (see Fig. 1). Fig. 10(a) shows the green power recorded during a period of more than 8 hours (30000 seconds). A stable SH power is obtained with a standard deviation of 0.011mW. For comparison, inset in Fig. 10(b) shows the green power recorded at the output of crystal C during 50 minutes (3000 seconds) without stabilization. A drift of 0.084 mW/h is observed with fast oscillations due to the air-conditioning system of the room.

**6. Conclusions**

We have investigated experimentally the single-pass frequency doubling of the 1029nm radiation from an Yb:fiber laser using bulk and waveguide PPLN crystals. While for the bulk crystal, only 4mW of green light were obtained from 400mW of IR, the waveguide crystals have high values of normalized conversion efficiencies ($278\% \cdot W^{-1} \cdot cm^{-2}$ and $613\% \cdot W^{-1} \cdot cm^{-2}$) and for one of them, a power of 33mW of green light was achieved with 197mW of incident IR power. Those results are limited by the coupling efficiency of the input IR power into the guide (63% and 45%) which could be enhanced by direct fibre injection.

We carried out the detailed characterisation of the waveguide crystals. The SHG process appears to be affected by geometrical inhomogeneities of the waveguide and a slight saturation for high output powers was observed. A shift of the temperature control for QPM was observed and attributed to light absorption in the waveguide.

The stability in time of SHG process efficiency in waveguide crystals was also studied. In the short term, a decrease of the SH power was observed after a few hours for high IR input power coupled in the guide. This decrease was reversible and the initial efficiency is recovered if the crystal is not irradiated for a few days. Nevertheless, over the long term, the conversion efficiency has been progressively reduced in a permanent way and after more than 2 years of use, it corresponds to only 60% of its initial one.

**Acknowledgments**

We are indebted to the electronics staff for their technical support, and fruitful discussions. This work is supported by the Mairie de Paris (program Research in Paris 2010), LNE, l'Observatoire de Paris and Action spécifique GRAM of the CNRS.
One of us (J.H.) acknowledges the Grant Agency of the Czech Republic (project GPP102/11/P820).

**Figure captions**

Fig. 1: Experimental setup: PBS (polarizing beam splitter), AOM (acousto-optic modulator), PD (photodiode), L (lenses).

Fig. 2: Photo of the temperature-controlled oven for PPLN crystal.

Fig. 3: Temperature stabilization for crystal C.
Record of the stabilized temperature during 40 hours.

Fig. 4: Characteristics of the bulk crystal.
(a) SH power vs squared IR input power collected at 31°C and 1029.351nm. (b) SH power vs temperature collected for 400mW of input IR power. (c) SH power vs input wavelength collected at 31 °C and 400mW of input IR power.

Fig. 5: SH power versus IR input power collected at fixed crystal temperature and laser wavelength (crystal B: 58.42°C and 1029.321nm; crystal C: 66°C and 1029.322 nm). Blue squares: crystal B; red circles: crystal C. The filled symbols correspond to measurements operated at the beginning of the use of the crystals. The empty symbols correspond to measurements operated after 2 years of work. Solid lines: fits according to eq. 1. Dash line: interpolation for data of crystal B at the beginning of the use.

Fig. 6: Optical conversion efficiency versus the IR input power.
squares: crystal B; circles: crystal C. The filled symbols correspond to measurements operated at the beginning of the use of the crystals. The empty symbols correspond to measurements operated after 2 years of work. Solid lines: fits according to theory. Dash line: interpolation for data of crystal B at the beginning of the use.
New measures: SH power versus IR input power collected at fixed crystal temperature and laser wavelength. For crystal B, the data are collected at 58.42°C and 1029.321 nm; crystal C, at 1029.322 nm and 66°C. new measures: optical conversion efficiency for crystal B and C, collected at fixed wavelength (1029.161 nm) and optimizing for each value the IR power coupling and the temperature.

Fig. 7: Wavelength dependence of the SHG power.
Crystal B: temperature 58.56°C; input IR power 31mW (coupled power 19.5mW). Crystal C: temperature 59°C; input IR power 100mW (coupled power 45mW).

Fig. 8: Temperature dependence of the SHG power.
Crystal B: IR wavelength 1029.320nm; input IR power 31mW (19.5mW coupled IR power). Crystal C: IR wavelength 1028.835 nm; IR power 100mW (coupled power 45mW).

Fig. 9: Evolution of the QPM temperature with the coupled IR power for crystal B.
SH power versus the temperature of the crystal at different coupled IR power. Inset: QPM temperature as a function of the coupled IR power.

Fig. 10: SH power collected at the output of crystal C for 60mW of IR input power with stabilization.
Inset: SH power collected in the same conditions without stabilization.

**Figures**

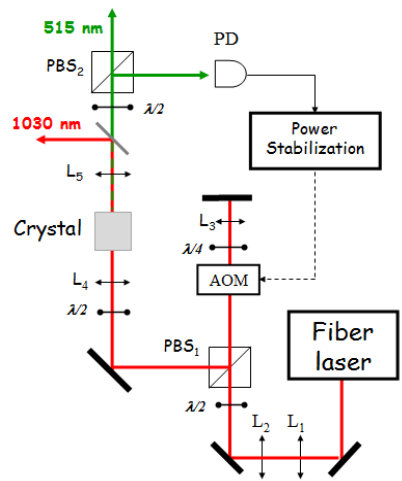

Figure 1.

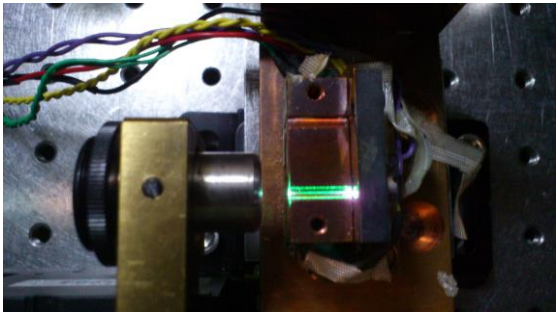

Figure 2.

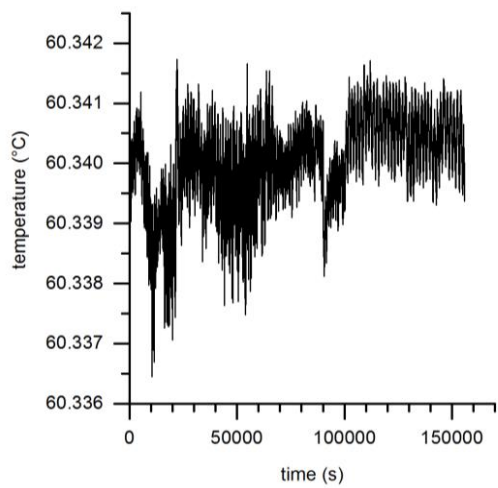

Figure 3.

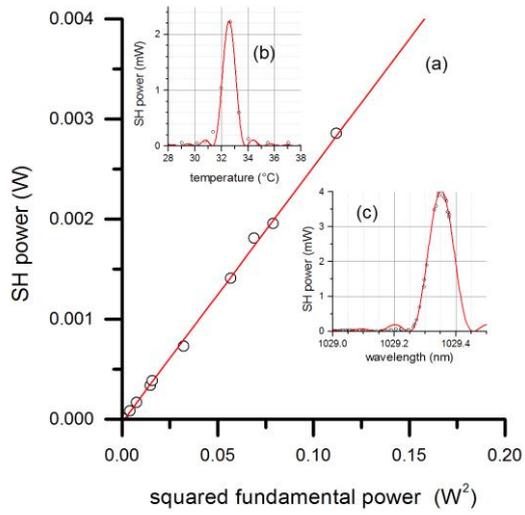

Figure 4.

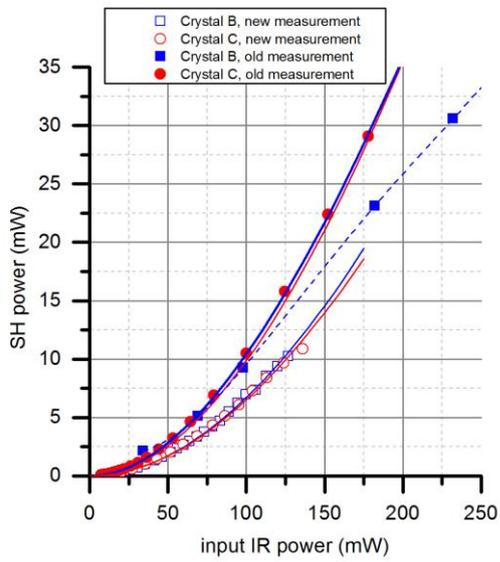

Figure 5.

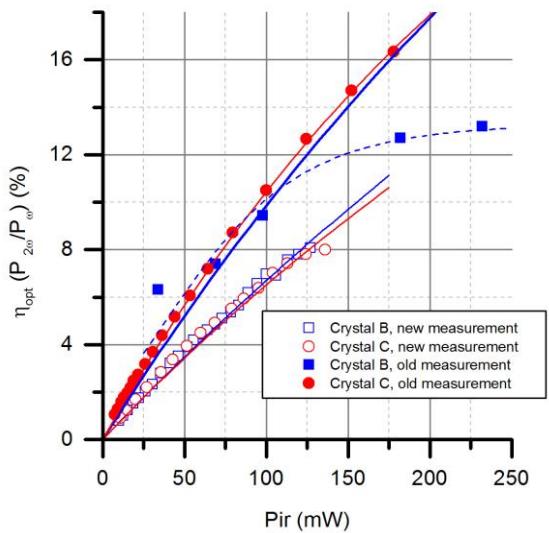

Figure 6

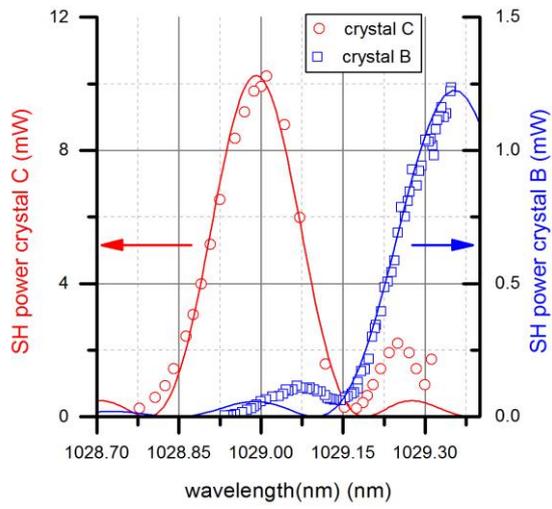

Figure 7.

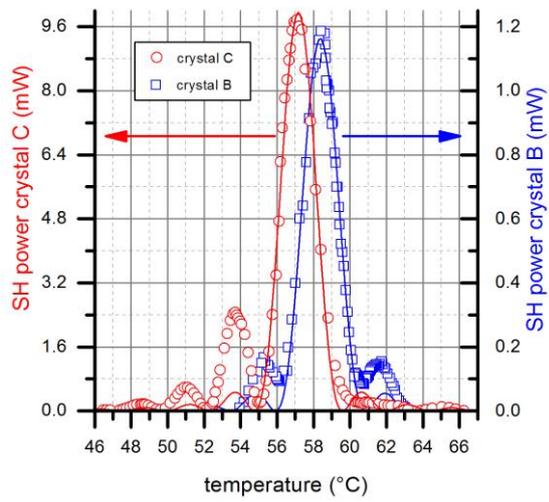

Figure 8.

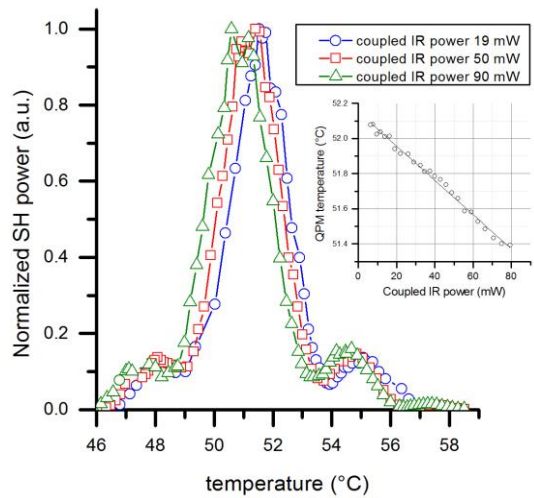

Figure 9.

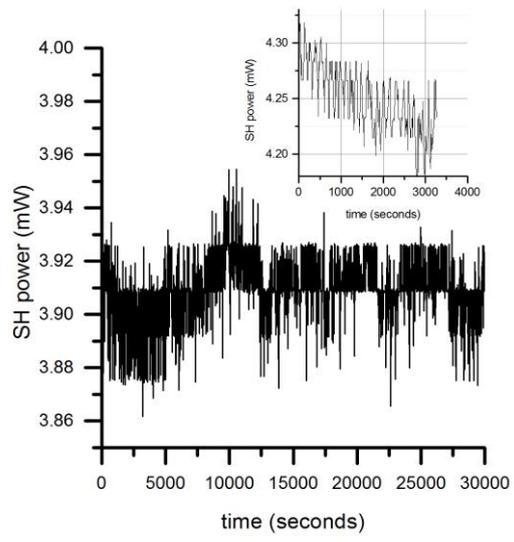

Figure 10.